# High-Resolution Electron Paramagnetic Resonance

Colin J. Stephen[1], Anton Tcholakov[1], Maik Icker[1], Stuart M. Graham[1], Xiaoming Zhao[2], Robert Day[1], Jeanette Chattaway[1], T. John S. Dennis[2], Wolfgang Harneit[3] and Gavin W. Morley[1]

1 Department of Physics, University of Warwick, Gibbet Hill Road, Coventry CV4 7AL, UK

2 School of Physics and Astronomy, Queen Mary University of London, Mile End Road, London, E1 4NS, UK

3 Department of Physics, Universität Osnabrück, Osnabrück 49076, Germany

Abstract: Electron paramagnetic resonance (EPR) is a valuable tool for physics, chemistry, biology and medicine, providing complementary spectroscopic information to NMR. It has long been known that EPR at high magnetic fields offers greater spectral resolution, but limitations in the THz instrumentation have prevented the full realisation of these opportunities. Here we describe an EPR spectrometer at the high magnetic field of 14 T using 396 GHz excitation, which adapts techniques from liquid-state NMR to obtain sharp EPR resonances with a width of 210 ppb (full-width half-maximum). We use this to measure resonance positions, and hence g-factors, with a precision that reaches ±16 ppb. Our use of in-situ liquid-state NMR of our solvent within the same sample improves the accuracy of these measurements: it allows us to reference our EPR measurement back to dilute gas $^3$He NMR for which quantum calculations are accurate. We measure the g-factor of N@$C_{60}$ in deuterated toluene as g = 2.002 099 09 (3), where the 3 in brackets means that the uncertainty on the last digit is ±3.

Resonances which can be resolved can be studied, and the high resolution of liquid-state NMR has made it indispensable for analytical chemistry. Previous high-field EPR [1-10] has sometimes been combined with NMR [11, 12], particularly for DNP studies [13-18]. The cyclotron resonance of a single trapped electron allowed a measurement of g = 2.002 319 304 361 70 (152): better than 1 ppt [19]. This measurement agrees with quantum mechanical calculations. EPR measurements have not come close to this level of precision or accuracy.

The most precise EPR g-factor we are aware of has been measured for metallic lithium in LiF as 2.002 293(2) with an uncertainty of 1 ppm [20]. The proton NMR of water was used to achieve high accuracy. Unfortunately, different samples of LiF:Li are known to have different g-factors at this level of precision, which may result from different impurities in them. The g-factor of carbon fibres was measured as 2.002 644(30), where a relative uncertainty of 15 ppm was achieved using the Overhauser effect as an in-built nuclear spin reference to get high accuracy [21]. For N@$C_{60}$ molecules rotating in the solid state, an EPR measurement of g = 2.002 04(4) was reported at ambient temperature at 5 T [12] with 20 ppm uncertainty. The accuracy came from in-situ referencing to liquid-state NMR of protons. The full-width half-maximum (FWHM) of the resonance was 0.46 MHz, which is 16 µT or 1750 ppb. An EPR g-factor has been measured with 25 ppb precision for $K_3CrO_8$ in $K_3NbO_8$; the magnetic field was measured by putting an NMR Gaussmeter in the sample position after removing the sample [22].

The sharpest EPR resonance reported to date, to our knowledge, is 400 ppb: this was for electron bubbles in superfluid liquid helium where a resonance was measured with a width of 0.2 µT at 13.5 GHz [23, 24]. At 9.67 GHz, a linewidth of 0.3 µT (900 ppb) was recorded for N@$C_{60}$ dissolved in $CS_2$ [25] making use of an unusually homogeneous electromagnet. The g-factor reported was 2.0036. Other N@$C_{60}$ papers have reported the g-factor as 2.00212(5) [26], 2.0022(3) [27], 2.00211(4) [28] and, in the paper that discovered this molecule, 2.0030(2) [29].

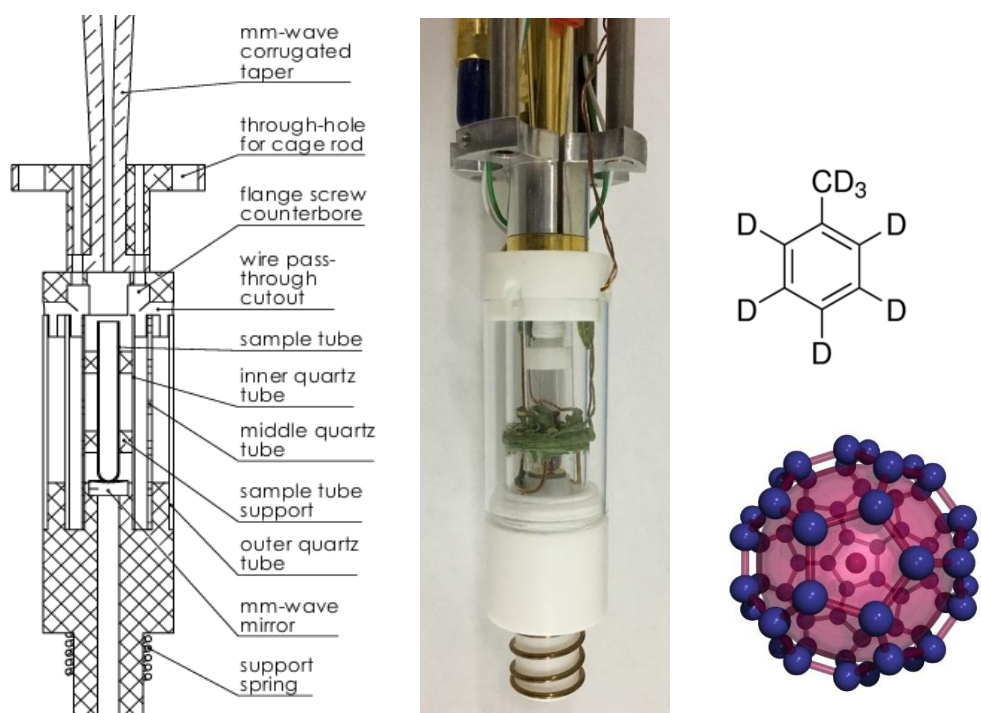

Figure 1. Sample probe for high-resolution electron paramagnetic resonance and NMR: schematic and photo. The structures of fully deuterated toluene and N@$C_{60}$ are shown also.

The key ingredients of our measurement are a high magnetic field from a persistent-mode superconducting NMR magnet with high homogeneity, and molecules tumbling fast enough to average out their environment. The high homogeneity requires symmetrically wound superconducting coils with shim coils, combined with an appropriate sample probe. We have built a probe that combines the best parts of liquid-state NMR and liquid-state high-field EPR with one sample tube. The probe was designed to be cylindrically symmetric along the direction of the magnet bore, using high-purity silica because of its low magnetism. It is not convenient to maintain this symmetry for the RF sample coil or the glue to attach it, so these are made from diamagnetically compensated materials: alloys or mixtures chosen so that their weak diamagnetism and weak paramagnetism mostly cancel. While EPR sample probes are generally based around microwave resonators, it becomes more difficult to build these for frequencies above 90 GHz because the wavelength becomes smaller than convenient machining tolerances [1]. When there is not a shortage of sample, a similar sensitivity is achieved with 'bucket resonators': almost non-resonant probes with sample volumes much larger than the wavelength [2]. Even when these bucket resonators are used, the magnetic field is normally swept to obtain an EPR spectrum: it is very unusual to sweep the microwave frequency [30]. Sweeping the magnetic field causes hysteresis which limits the precision and accuracy of measurements; a superconducting magnet should be left in persistent mode for over a week before it stops drifting enough for high-resolution NMR. We use a bucket resonator and sweep the microwave frequency to obtain a spectrum. The curved mirror below the sample to reflect the 396 GHz back out of the probe is thinner than the skin depth of the NMR radiofrequency to avoid disturbing it. We use standard magnetic field modulation for our lock-in amplifier.

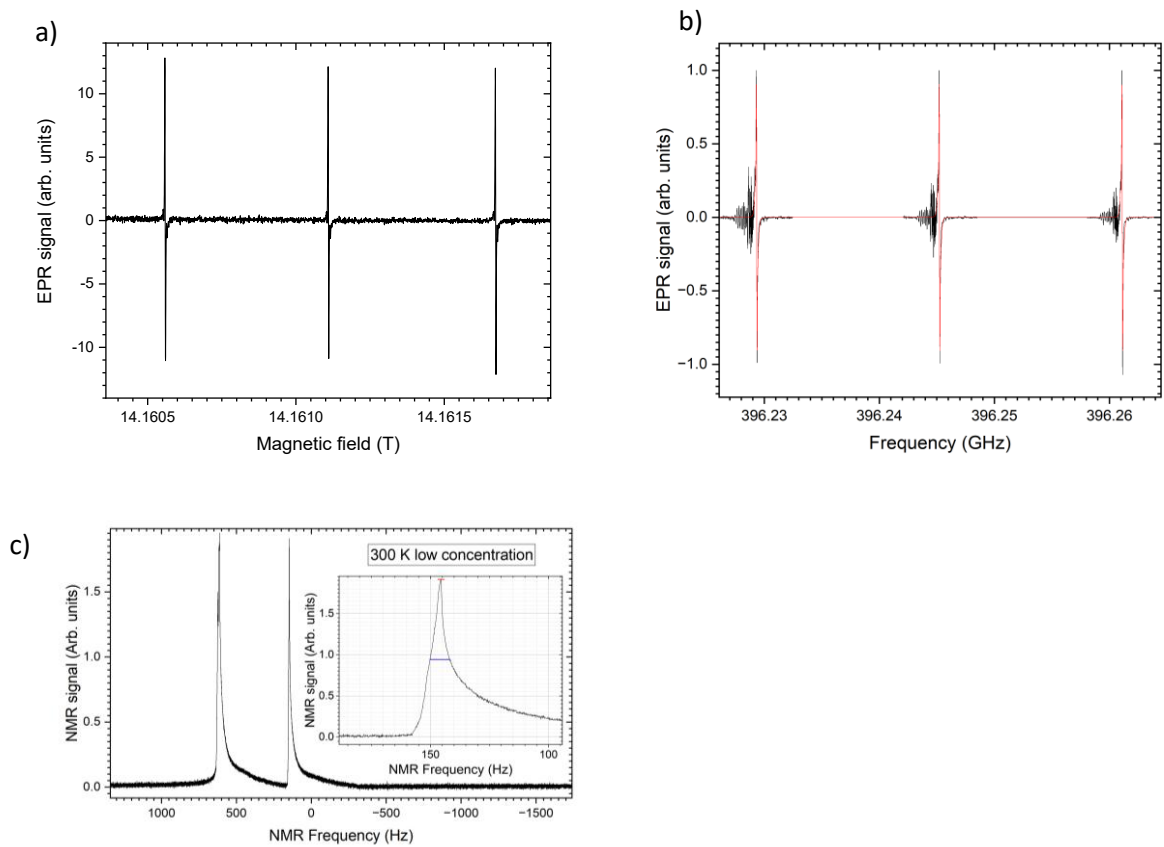

Figure 2. Electron paramagnetic resonance spectra of N@C$_{60}$ at 300 K temperature using a) magnetic field sweep and b) THz frequency sweep of a different sample. c) $^2$H NMR of the fully deuterated toluene solvent taken immediately after recording the central EPR resonance in spectrum b); a zoom of the NMR resonance from the CD$_3$ group is also shown, with a blue horizontal line showing the full-width half-maximum of 8.4 Hz, and a red horizontal line showing the estimated uncertainty in the peak of ±1.4 Hz.

To ensure that the EPR detection system is suitable for high resolution, we built a homodyne 396 GHz EPR bridge with a low-phase-noise source and a 12.4 GHz frequency counter that was calibrated less than a month before the measurements in the main paper. The free-space quasi-optic mirrors were made by Thomas Keating Ltd.. The 396 GHz source starts with a 12.4 GHz source (Agilent E8257D) which is upconverted by a factor of 32 by an amplifier/ multiplier chain (Virginia Diodes Inc). The 12.4 GHz source and the 12.4 GHz frequency counter were placed far enough outside the stray field of the magnet so they were in a magnetic field of less than 300 µT. A heterodyne bridge would introduce phase noise between the two THz frequencies so is avoided. The EPR resonances of N@C$_{60}$ in solution that we measure are around 3 µT which is close to the 0.3 µT linewidth of previous low-field N@C$_{60}$ measurements [25], but by going from 10 GHz to 396 GHz we improve the linewidth from 900 ppb to 210 ppb. This small linewidth gives us high precision. Our superconducting magnet has been left in persistent mode for years before these measurements. We shim it with in-situ liquid-state NMR of our solvent. To obtain an accurate EPR g-factor, we use our in-situ liquid-state $^2$H NMR of the deuterated toluene solvent with previously reported measurements referencing $^2$H NMR of deuterated toluene to dilute $^3$He gas NMR and quantum calculations of isolated $^3$He gas NMR [31].

We collect both magnetic-field swept and THz-frequency-swept spectra. The latter are more accurate and precise because they avoid magnet hysteresis. We simply compare the frequency of the EPR and the NMR at the same magnetic field, without needing to calculate the magnetic field. The frequency-swept EPR spectra show unwanted resonances due to sweeping through the resonance of low-Q cavities formed along the THz beam path. The magnetic-field swept spectra do not have these features showing that they are not EPR features.

The N@$C_{60}$ molecule is a single nitrogen atom trapped in the centre of a $C_{60}$ fullerene cage. There is an electron spin of S = 3/2 hyperfine coupled to a $^{14}$N nuclear spin of I = 1. The spin Hamiltonian is

$$\mathcal{H} = g\,\mu_B B_z S_z - g_N\,\mu_N B_z I_z + A I \cdot S + D\left[S_z^2 - \frac{1}{3}S(S+1)\right] + E(S_x^2 - S_y^2). \qquad (1)$$

From a fit to the frequency-swept EPR in Figure 2 b) we obtain the central frequency of the EPR as $f_{EPR}$ = 396,245,239,000 Hz with an uncertainty of 4 ppb. We wrote our own software for fitting the frequency-swept EPR to Equation 1. The fitted value of the hyperfine coupling constant is A = 15.8929 MHz ± 0.3 kHz. The three EPR resonances in Figure 2b were collected separately as they each needed different positions of the moving mirror which is used to provide the local oscillator (LO) to the 396 GHz detector. This moving mirror is a PTFE plate which reflects some of the 396 GHz excitation before it goes to the sample probe. Moving it up and down controls the phase of the LO with respect to the signal that comes from the sample probe. This is set to have a differential EPR lineshape. The different wavelengths for the three different EPR resonances mean that the phase of the LO with respect to the signal is different if the moving mirror is not adjusted. The 396 GHz power from the source was also different at different frequencies, and along with the different setting of the moving mirror, this led to different 396 GHz power being detected. This was compensated by changing the Martin Puplett interferometer which changes the polarization of the 396 GHz sent to the probe. Linear polarization does not make it to the detector, so partially circular polarization was used to provide some LO to the detector which increases the signal-to-noise. The amplitude of the three frequency-swept resonances cannot therefore be compared: they were all baselined and normalised. The field-swept spectrum shows that the three resonances are all of about equal intensity.

The NMR spectrum taken immediately after the middle EPR resonance is shown in Figure 2 c). The shimming of the magnet was not good enough to get a Lorentzian or Gaussian resonance: there is a shoulder to the right of the graph. The full-width half-maximum (FWHM) of the NMR is 8.4 Hz which is 91 ppb: this is sharper than the EPR resonance which is has a FWHM of 210 ppb. The EPR lineshape will be a convolution of the NMR lineshape and other things such as modulation broadening, power broadening and the intrinsic linewidth determined by $1/T_2^*$, where $T_2^*$ is the inhomogeneous spin dephasing time. As the NMR is sharper than the measured EPR, the non-ideal magnet shimming is not a significant source of broadening for the EPR. We therefore take the NMR resonance frequency as being that of the peak of the data rather than fitting it with a function and taking the peak of the function as the NMR resonance frequency. We estimate the uncertainty in this NMR frequency as being 8.4 Hz/6 = 1.4 Hz, which is 15 ppb. The NMR resonance peak for the $CD_3$ group in the toluene-d8 is 145.6 Hz above the spectrometer transmit frequency of 92,418,431 Hz so $f_{NMR}$ = 92,418,486.6 Hz.

These EPR and NMR measurements were made while holding the temperature at 300 K using a Cernox$^{TM}$ thermometer in the cryostat and a Lakeshore temperature controller. The temperature reading was stable to ±20 mK.

To calculate the EPR g-factor ($g$) from our measurements, we use the resonance conditions for EPR and NMR:

$$h f_{EPR} = g \mu_B B, \quad (2)$$

$$h f_{NMR} = (1-\sigma) \mu_{2H} B, \quad (3)$$

where $h$ is Planck's constant, $\mu_B$ is the Bohr magneton, $\mu_{2H}$ is the nuclear magnetic moment of deuterium, $B$ is the magnetic field and $\sigma$ is the absolute shielding constant for the deuterium nuclear spins in the $CD_3$ group in fully deuterated toluene (toluene-d8). Dividing equation 2 by equation 3 and rearranging we get:

$$g = (1 - \sigma) \frac{\mu_{2H}}{\mu_B} \frac{f_{EPR}}{f_{NMR}}. \quad (4)$$

The deuteron magnetic moment and the Bohr magneton are reported in CODATA 2022 as $\mu_{2H}$ = 4.330,735,087 × $10^{-27}$ J/T ± 2.6 ppb and $\mu_B$ = 9.274,010,065,7 × $10^{-24}$ J/T ± 0.31 ppb. The shielding constant for toluene-d8 has been measured previously at 300 K, by doing NMR of it along with dilute $^3$He gas [32]. Combining this with relatively simple quantum calculations of the NMR of $^3$He gas provides $\sigma$ = 31.525 ppm [32]. We therefore reach a g-factor for N@$C_{60}$ of 2.002 099 09 (3). The uncertainty in the g-factor is dominated by the 15-ppb uncertainty in the NMR resonance frequency.

We measure the concentration of fullerenes in this 'dilute' sample as 0.66 mg/ml using UV-vis spectroscopy to compare to a known sample. This is the mass of the N@$C_{60}$ plus the mass of the $C_{60}$ divided by the volume of deuterated toluene. We measure the doping ratio as 1.1% using high-performance liquid chromatography (HPLC) and EPR. This is the ratio of the number of N@$C_{60}$ molecules to the sum of the N@$C_{60}$ and $C_{60}$ molecules.

We used the same technique to measure a sample that is 3.5 times more concentrated in terms of the mass of fullerenes per volume of solvent, at both 300 K and 291 K. The EPR linewidth of the more concentrated sample at 300 K was slightly higher, and reducing the temperature of this sample reduced the EPR linewidth slightly, as shown in Figure A5c of the Appendix. For the high-concentration sample at 300 K, the magnet shimming was the same as for the dilute sample at 300 K, and although the sample probe was taken out of the magnet to change the sample, putting it back into the magnet produced a $^2$H NMR spectrum with the $CD_3$ peak at the same frequency of 92,418,486.6 Hz. The fit to the EPR produced a resonant frequency of 396.245 258 07 GHz ±4.6 ppb. This is a slightly higher g-factor than the dilute sample at the same temperature.

For measurements of the concentrated sample at 291 K, the magnet shimming was not as good simply because this measurement was taken before the shimming was improved for the 300 K measurements. The NMR lineshape was not a Lorentzian or Gaussian as shown in the Appendix. The NMR peak was at 92,417,639.5 Hz with a FWHM of 25.7 Hz (278 ppb). The NMR uncertainty was estimated as FWHM/6 = 46 ppb. The EPR resonance from the fit was at 396.241 681 79 GHz ±1 ppb. Figure 3 shows the three frequency-swept g-factors measured here.

| | Temperature (K) | Concentration (mg/ml) | g-factor | Relative g-factor uncertainty (ppb) |
|---|---|---|---|---|
| Dilute sample | 300 | 0.664 ±0.05 | 2.002 099 09 (3) | 16 |
| Concentrated sample | 300 | 2.3 ±0.35 | 2.002 099 19 (3) | 16 |
| Concentrated sample | 291 | 2.3 ±0.35 | 2.002 099 47 (9) | 46 |

Table 1. Measurements of the g-factor for N@$C_{60}$ in toluene-d8 from this work, shown in Figure 3.

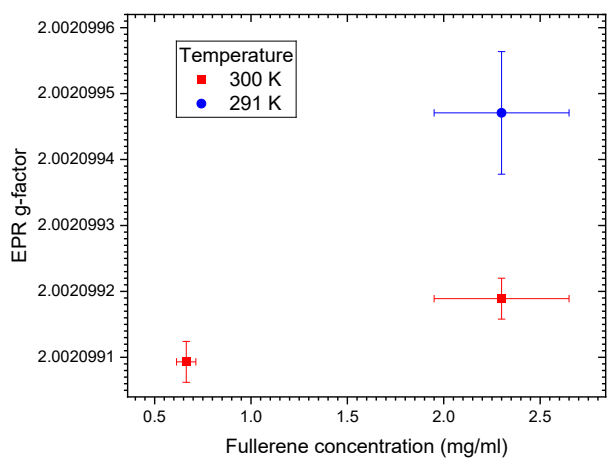

Figure 3. Measured g-factors for N@$C_{60}$ at two temperatures and for two sample concentrations, using frequency-swept EPR at 14 T.

To estimate the relative uncertainties on the g-factors (in ppb) the relative uncertainties on $f_{EPR}$, $f_{NMR}$, $\sigma$, $\mu_{2H}$ and $\mu_B$ were combined in quadrature. However, the uncertainties on the latter three quantities are 1 ppb [32], 2 ppb (CODATA 2022) and 0.31 ppb (CODATA 2022), which are so small that they are not significant here. The uncertainties in the g-factors measured here are limited by $f_{NMR}$ which is limited by the quality of our magnet shimming.

The EPR of lithium metal particles in LiF was measured with a field sweep at 14 T, but the spectrum was split into multiple peaks, presumably due to the different shapes and sizes of the particles. This splitting is not normally detectable at the low magnetic fields typically used for EPR.

These results will spur the development of calculations of high-resolution g-factors and hyperfine coupling constants, in addition to calculations and simulations of how these are shifted by the use of different solvents, temperatures and pressures. The use of this spectrometer with EPR of electron bubbles in superfluid liquid helium [23, 24], combined with in-situ NMR of the dilute $^3$He may provide new opportunities for metrology as the EPR linewidth should be even sharper than that of N@$C_{60}$.

Extending the measurements presented here to nitrogen-vacancy (NV) centres in diamond would be valuable. Absolute NV magnetometry[33] and tests of the quantum nature of gravity [34] will require more accurate and precise knowledge of the NV g-factor and zero-field splitting which are only known to 5 and 4 significant figures respectively [35, 36].

Acknowledgements

Thanks to Oleg Peter and Petra Bösel (Osnabrück University) for assisting with the preparation of the N@$C_{60}$ sample. Thanks to the Warwick Solid-State NMR Group and in particular Andy Howes, Dinu Iuga, Tom Kemp, Ray Dupree R.I.P., Kevin Pike, Wing Chow and Michael Hope.

Appendix

Figure A1 shows a schematic and a photograph of the THz bridge. Figure A2 shows the data and protocol used to collect a field-swept EPR spectrum with NMR referencing.

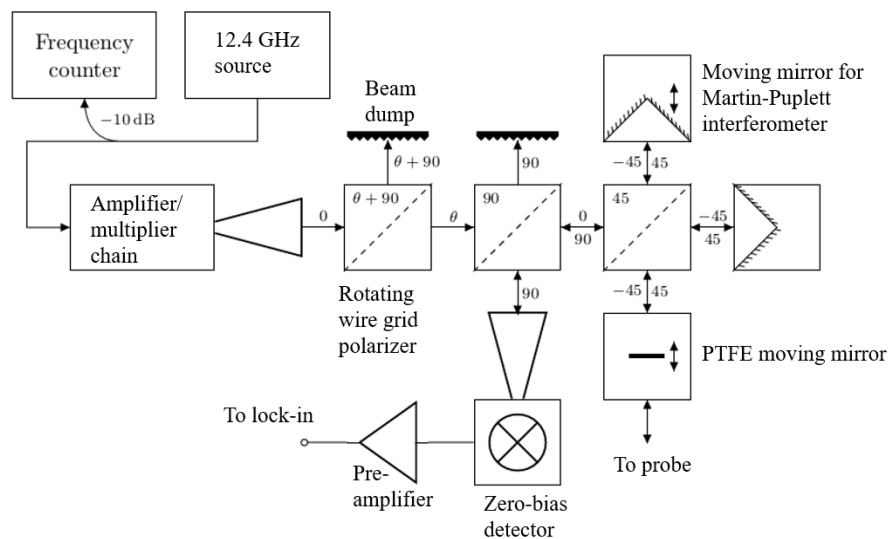

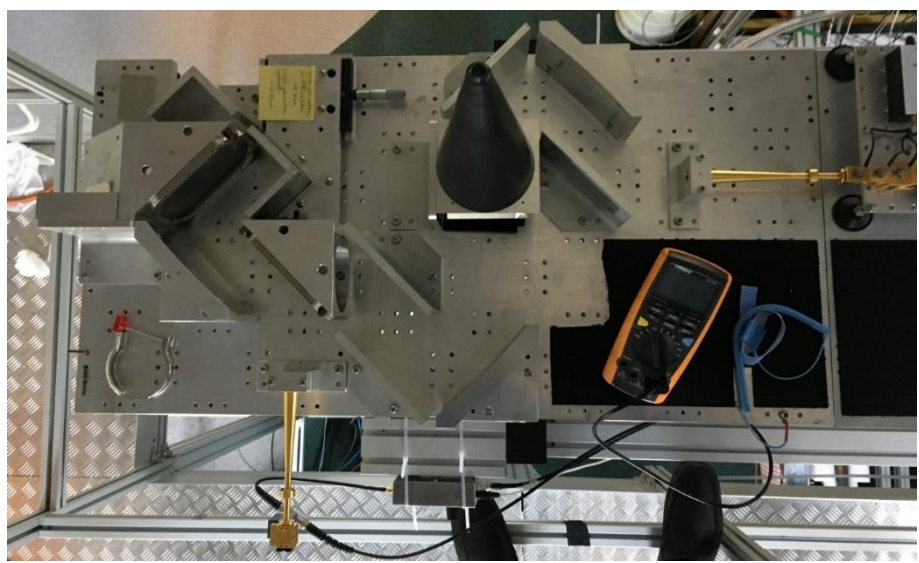

Figure A1. The quasi-optic free-space 396 GHz bridge: schematic and photograph. The amplifier/ multiplier chain which multiplies the input microwave frequency of around 12.4 GHz by 32. A Martin-Puplett interferometer is used to control the polarization of the 396 GHz that gets sent to the probe. A rotating polarizing wire grid is used to control the power of the 396 GHz excitation by dumping some of the power from the source which is kept constant. A zero-bias detector made by VDI is used to detect the 396 GHz signal. A moving mirror (not shown in the photo) between the bridge and the probe is used to control the phase of the 396 GHz local oscillator with respect to the 396 GHz signal that returns to the detector from the sample probe.

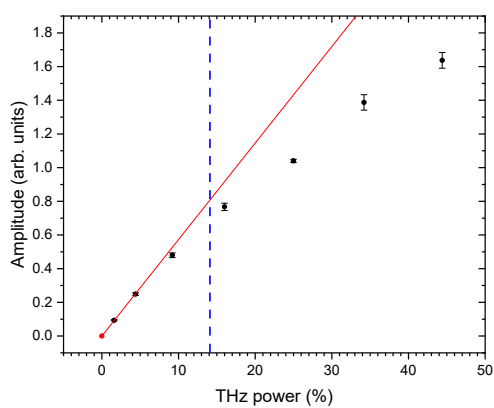

Figure A2. Fitted amplitude of frequency-swept EPR at 291 K for different 0.396 THz power. The red point at 0% was not measured but we assume that there would be no amplitude at zero THz power. The red line is a linear fit to the three points at 0%, 1.6% and 4.4%. The dashed blue line shows the 14% power used in the main paper which was a compromise between having good signal-to-noise and avoiding saturation.

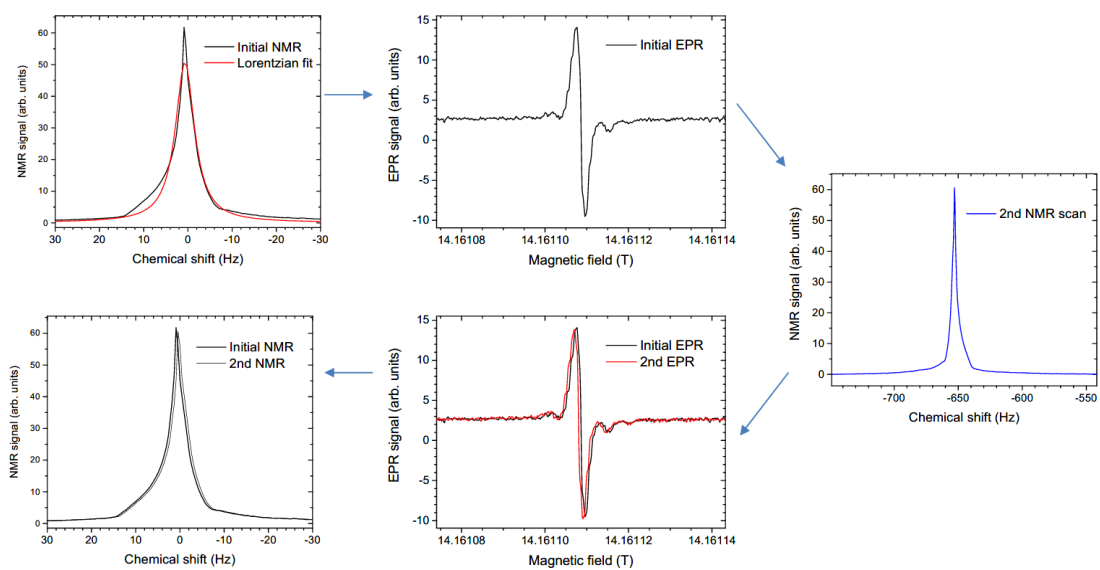

Figure A3. Protocol for magnetic-field-swept EPR: NMR is done before and after an EPR field sweep, and then again after measuring the EPR again by sweeping the field back down. This gives EPR spectra for the up and down sweeps with the magnetic field at the beginning and end of each determined by NMR. The up- and down-swept EPR spectra are slightly offset due to magnet hysteresis so they are shifted towards each other to make a calibrated spectrum. The EPR frequency sweeps in the main paper have no magnet hysteresis. The NMR spectra in this figure are better shimmed due to using a different set of room-temperature shim coils which are no longer available.

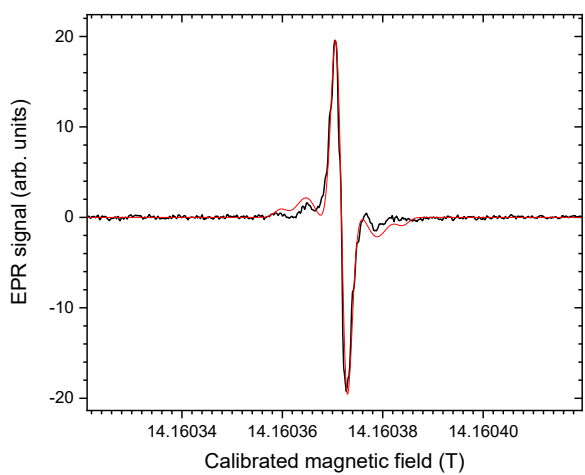

Figure A4. Field-swept EPR spectrum of one N@C$_{60}$ resonance, fit with Easyspin software. The magnetic field has been calibrated using NMR before, during and after an up-sweep of the magnetic field and a down-sweep.

Figure A4 is a field-swept EPR spectrum showing one resonance fit with Easyspin [37] using a zero-field splitting (ZFS) of D = 170 kHz, a ZFS E strain of 35 kHz, and unresolved hyperfine coupling (H strain) of 80 kHz (Gaussian linewidth) in the x, y and z directions of the molecular frame. Anisotropic terms such as D are not expected for N@C60 tumbling in solution, but our use of a high EPR frequency means that anisotropies may not be fully averaged quickly enough. The unresolved hyperfine coupling could be due to $^{13}$C which has been resolved in X-band EPR [25]. The fitting parameters used are not unique: choosing other values improves some areas of the fit while making others worse. Figure A5a) shows the three N@C$_{60}$ hyperfine lines shifted in magnetic field to lie on top of each other: their lineshapes are very similar.

During the field-swept measurements, the 12.4 GHz microwave source and the frequency counter used to read it were placed too close to the superconducting magnet meaning they were in a stray field of around 2 mT. The source and counter use ferrimagnetic YIG (yttrium iron garnet) crystals; while these electronics boxes did not show a fault in the 2 mT magnetic field, it is possible that they were not performing as precisely as they should have. This may have caused a systematic error in the frequency, so the g-factor measurements taken from the field-swept data are not reliable. For the frequency-swept measurements in the main paper, this problem was solved by putting both microwave boxes over six meters from the magnet where the stray field was less than 300 µT. By reading out the microwave frequency on the calibrated counter to get the x-axis of the frequency-swept data, problems of frequency drift are automatically avoided.

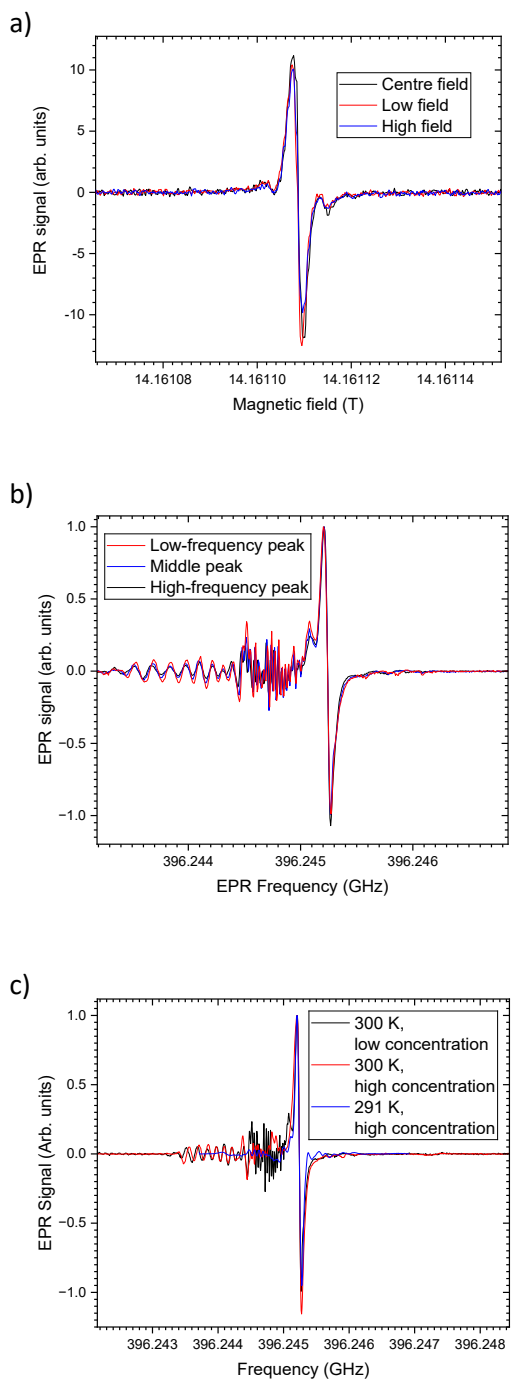

Figure A5. Shifting the magnetic field (a) or frequency (b) of the three N@C$_{60}$ resonances so they overlap shows their similar lineshapes. Spectrum a) is a field-swept measurement and b) is the frequency swept measurement from Figure 2b in the main paper. Both of these were with the dilute sample at 300 K. Almost all of the small oscillations in b) are due to unwanted 396 GHz cavities formed along the length of the experiment. c) Plotting the central EPR resonance for the two samples at two different temperatures, with the x-axis shifted to make them overlap. This shows that the small oscillations vary with sample temperature. The sample probe was taken out and re-inserted to change to the lower concentration sample: this will have changed the 396 GHz path, including a different liquid level. The EPR resonance is slightly sharper at 291 K than 300 K, and slightly sharper for the lower spin concentration compared to the higher spin concentration.

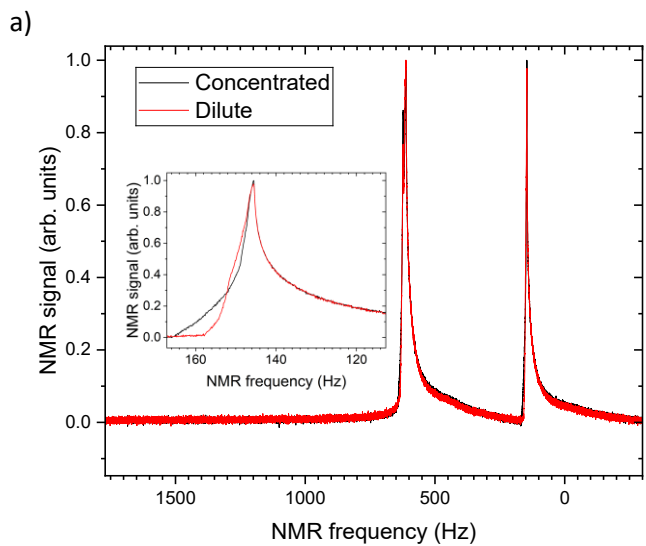

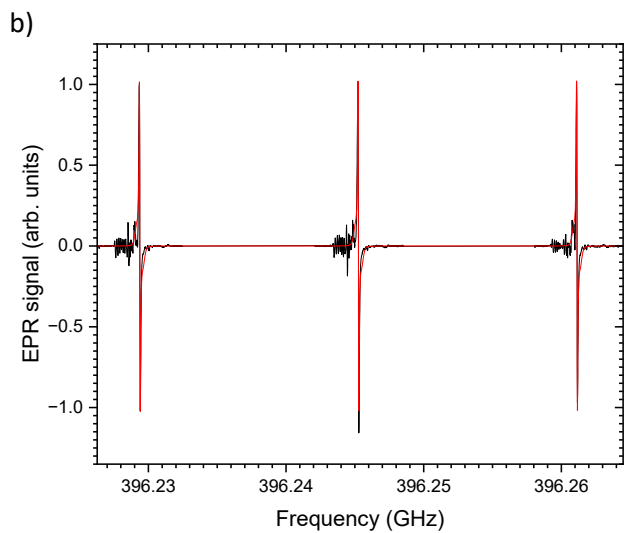

Figure A6. a) $^2$H NMR at 300 K of the toluene-d8 for the concentrated and dilute samples of N@C$_{60}$ for the frequency-swept EPR in the main paper. The magnet shims were not changed between these measurements, but the sample probe was taken out to change the sample. b) Frequency-swept EPR spectrum of the more concentrated sample of N@C$_{60}$ at 300 K.

The field-swept mode was used to measure the N@C$_{60}$ g-factor as a function of the concentration of the fullerene in the toluene-d8, but the data are not shown due to possible systematic errors in the microwave frequency due to the counter and source being in a stray field of about 2 mT. For these measurements, the error bars in the g-factor were larger and were dominated by the uncertainty in the EPR resonance position which was 19 to 57 ppb. This was larger due to uncertainty from magnet hysteresis, even though up and down sweeps were combined. The uncertainty in the NMR resonance position was only 0.2 ppb due to better shimming. Less than 1% of the C$_{60}$ cages contained nitrogen atoms for this sample. The measured g-factor varied by 25 ppb or less when the fullerene concentration was changed from 0.11 to 0.73 mg/ml with no clear trend; the uncertainties on these measurement were higher even ignoring the possible systematic error on the microwave frequency. Fullerenes cluster for these concentrations meaning the fullerenes are closer together than might otherwise be expected [38, 39]. High fullerene concentrations could also alter the NMR solvent shielding [40].

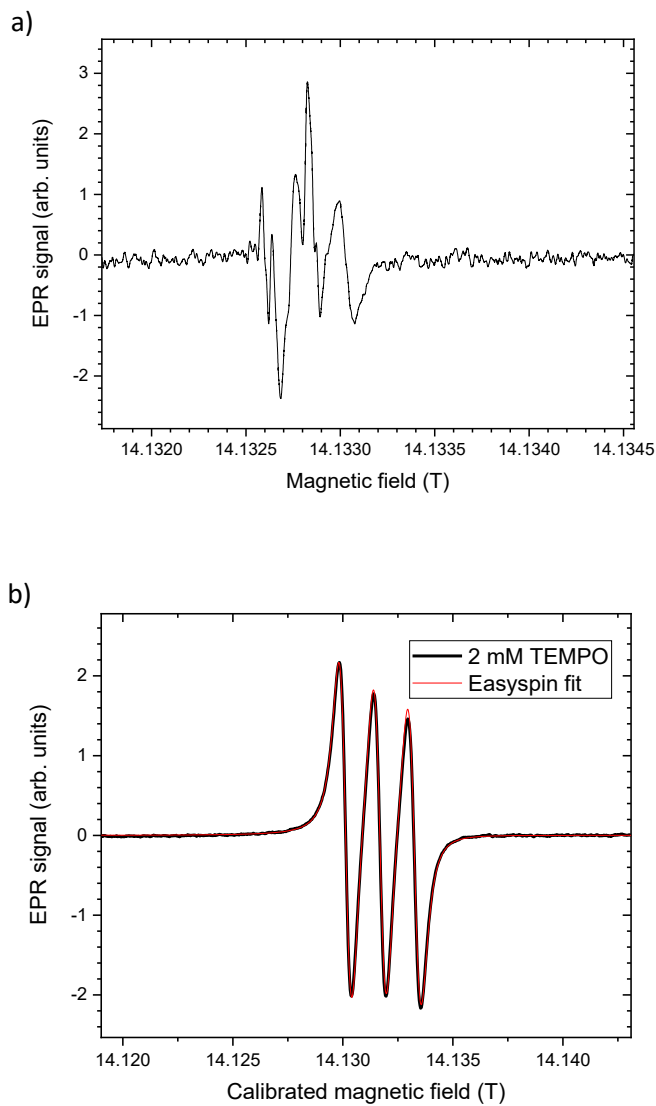

Figure A7. Field-swept EPR: a) Lithium metal particles in LiF shows splitting which is presumably due to the presence of particles of multiple sizes and shapes. b) TEMPO dissolved in toluene-d8, using $^2$H NMR to calibrate the magnetic field as shown in Figure A2. The data were fit with Easyspin [37] using a hyperfine constant of $[A_x, A_y, A_z]$ = [14 12 106] MHz, g-factors of
$[g_x, g_y, g_z]$ = [2.01043 2.00415 2.0039], rotational correlation time $t_{corr}$ = 6.5 ps, and linewidths of 0.35 mT Gaussian and 0.40 mT Lorentzian.

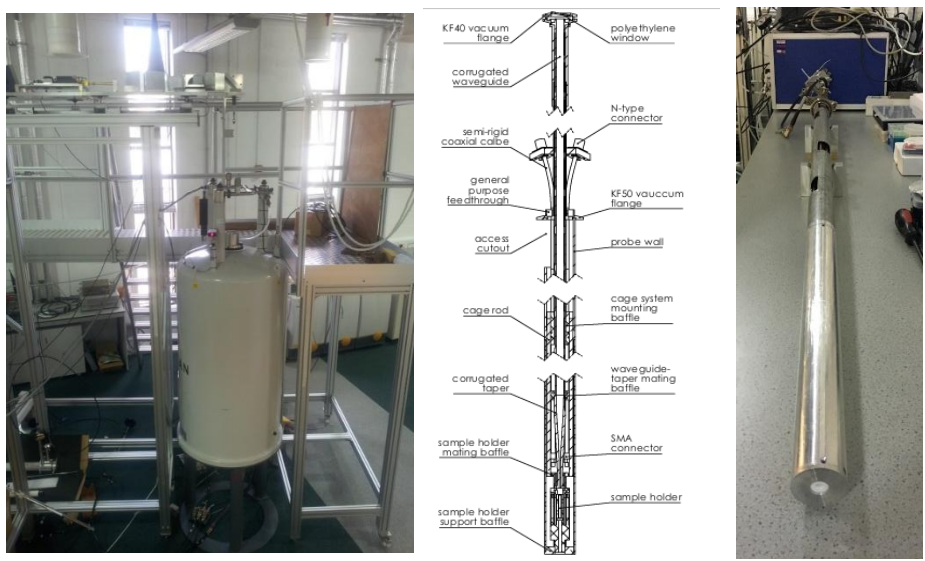

Figure A8. Photograph of the magnet, a schematic of the probe, and a photograph of the probe which is 2.2 m long with an overmoded corrugated waveguide inside it.